\documentclass[english,twoside,a4paper,10pt]{article}

\usepackage[latin1]{inputenc}
\usepackage[T1]{fontenc}
\usepackage{amsmath}
\usepackage{amsfonts}
\usepackage{graphicx}
\usepackage{subfigure}
\usepackage{a4wide}
\usepackage{amssymb}
\usepackage{fancyhdr}
\usepackage{mathrsfs}
\usepackage[toc,page]{appendix}

\begin{document}

\title{\bf Coupled fluids model in FRW space-time}
\author{ 
Shynaray Myrzakul\footnote{Email: shynaray1981@gmail.com},\,\,\,
Ratbay Myrzakulov\footnote{Email: rmyrzakulov@gmail.com},\,\,\,
Lorenzo Sebastiani\footnote{E-mail address: l.sebastiani@science.unitn.it
}\\
\\
\begin{small}
Eurasian International Center for Theoretical Physics and  Department of General
\end{small}\\
\begin{small} 
Theoretical Physics, Eurasian National University, Astana 010008, Kazakhstan
\end{small}\\
}

\date{}

\maketitle


\begin{abstract}
In this paper, we
analyze a two coupled fluids model by investigating several solutions for accelerated universe in flat FRW space-time. One of the fluids can be identified with the matter and the model possesses the standard matter solution also. Beyond the removal of the coincidence problem,
we will see how the coupling may change the description of the energy contents of the universe and which features can be aquired with respect to the standard decoupled cases.
\end{abstract}



\tableofcontents
\section{Introduction}

The dark energy issue, namely the accelerated expansion of the universe today~\cite{WMAP}, and other 
issues related to the existence of an early-time acceleration (the inflation) after the Big Bang, suggest
the presence in our universe of `dark' fluids different to standard matter and radiation. 
In fact, the simplest way to describe the cosmic acceleration in agreement with the observations is given by the introduction of small and positive
Cosmological Constant in the framework of General Relativity (the so called $\Lambda$CDM Model), but several descriptions are allowed. The Cosmological Constant represents the energy density of a perfect dark fluid whose Equation of State paramter $\omega$ is equal to minus one, namely the pressure is negative and induces the acceleration. Apart from the fact that the cosmological data  constrain $\omega$ to be very close to minus one without exluding different
forms of dark perfect fluid (quintessence, phantom...) with some varieties of future scenarios, there are other possibilities. For example, the dark components of the universe, whose origin remains unknown, may be not represented by perfect fluids, like for the large scale structure of standard matter (see Refs.~\cite{fluidsOd2, Alessia, Jamil} for inhomogeneous and viscous fluids and Ref.~\cite{B1} for some application of fluid cosmology to the inflation). An other possibility is given by the introduction of a coupling between the dark fluid and the matter. Such a possibility has been investigated in the past
in an attempt to solve the coincidence problem~\cite{Odintsovcouplingfluids,Viscousfluids}:
why we observe  matter today and dark energy almost equal in amount is an open question in the standard cosmology, but the introduction of the coupling between matter and dark energy renders them dependent on each other and removes the problem.

In this paper we would like to 
analyze different kinds of dark fluid coupled with matter in flat Friedmann-Robertson-Walker space-time. We will start from a two-fluids model and we will separate the contributes to the Hubble parameter coming from the fluids. In fact, one of the fluids will be identify with matter by putting its pressure equal to zero. Thus, the Friedmann equations result to be easy to solve at the price to introduce a coupling between the fluids. In this way, we can aquire several descriptions for accelerated universe. Namely, we will use our representation to investigate the solution of $\Lambda$CDM Model, the quintessence and phantom solutions, and singular and bounce solutions, the last ones used in inflationary scenario. We will see how the coupling can change the description of the energy contents of the universe and which features can be aquired beyond the removal of the coincidence problem.

The paper is organized as follows. In Section {\bf 2}, we will present the two coupled fluids model. The first fluid, with constant Equation of State parameter $\omega$, is able to describe the matter, and the second one is a dark fluid that will be used to have acceleration. In Section {\bf 3}, some accelerated solutions are investigated by using our formalism. Namely, the $\Lambda$CDM Model solution for matter and dark energy, the quintessence and phantom solutions and singular and bounce solutions. In Section {\bf 4}, to complete the work, we will extend the formalism to the case where also the first fluid has not a constant Equation of State parameter $\omega$, namely we will present an example of coupling between dark fluids for phantom universe. 
Conclusions are given in Section {\bf 5}.

We use units of $k_{\mathrm{B}} = c = \hbar = 1$ and denote the
gravitational constant, $G_N$, by $\kappa^2\equiv 8 \pi G_{N}$, such that
$G_{N}^{-1/2} =M_{\mathrm{Pl}}$, $M_{\mathrm{Pl}} =1.2 \times 10^{19}$ GeV being the Planck mass.


\section{Formalism}

Let us consider the flat Friedmann-Robertson-Walker (FRW) metric, 
\begin{equation}
ds^{2}=-dt^{2}+a^{2}(t)d\mathbf{x}^{2}\,,
\end{equation}
$a(t)$ being the scale factor of the universe. In our analysis, we will work with a two-fluids model, whose Friedmann equations read
\begin{equation}
\frac{3H^2}{\kappa^2}=\rho_\text{1}+\rho_\text{2}\,,\quad
-\frac{1}{\kappa^2}\left(2\dot H+3 H^2\right)=p_\text{1}+p_\text{2}\,,
\label{EOMs}
\end{equation}
where $H=\dot a(t)/a(t)$ is the Hubble parameter and the dot denotes the derivative with respect to the cosmological time. The energy density and the pressure of the two fluids are given by $\rho_\text{1,2}$ and $p_\text{1,2}$.
The (total) conservation law is derived from Friedmann equations as
\begin{equation}
\dot\rho_\text{1}+\dot\rho_\text{2}+3H(\rho_\text{1}+p_\text{1})+3H(\rho_\text{2}+p_\text{2})=0\,.\label{CL}
\end{equation} 
For the first fluid we assume the following Equation of State (EoS),
\begin{equation}
p_\text{1}=\omega_\text{1}\rho_{1}\,,\quad\omega_1>-1\,,\label{fluidone}
\end{equation}
with $\omega_\text{1}$ constant in the non-phantom region. This choice follows from the fact that for our discussions we will often identify this fluid with standard matter ($\omega_1=0$) in the specific examples.
When the contribute of the fluid two vanishes, the solution of Friedmann equations driven from the fluid one is
\begin{equation}
H_\text{1}  (t)=\frac{2}{3(1+\omega_1)\,t}\,,\quad a_\text{1}  (t)=a_{\text{1}  (0)}\,t^{\frac{2}{3(1+\omega_1)}}\,,\label{HDM}
\end{equation}
where $a_{\text{1}  (0)}$ is an integration constant and the pedex `$\text{1}$' distinguishes this solution. For example, if $\omega_1=0$, we recover the matter dominated universe with Hubble parameter $H_1=2/(3t)$. Note that this solution results to be always for expanding universe ($H_1>0$) due to the choice $\omega_1>-1$. In such a case, from the conservation law we also get
\begin{equation}
\rho_{\text{1}}(t)=\rho_{\text{1}  (0)}a_\text{1}  (t)^{-3(1+\omega_1)}
\equiv
\frac{4}{3\kappa^2(1+\omega_1)^2\,t^2}
\,,\label{rhodm}
\end{equation}
where $\rho_{\text{1} (0)}$ is an other integration constant eventually related with $a_{\text{1}  (0)}$. 

Let us introduce now the fluid two,whose nature will be `dark' and whose Equation of State could be in any form, and 
a coupling between the two fluids. Our aim is to start from (\ref{HDM}) to reconstruct some cosmological model for coupled fluids. By decomposing the Hubble parameter as
\begin{equation}
H(t)=H_\text{1}(t)+\tilde H(t)\,,\quad a(t)=a_\text{1}(t)\tilde a(t)\,,\label{tH} 
\end{equation}
and by assuming that (\ref{rhodm}) is still valid, the conservation laws (on shell) for the two fluids result
\begin{eqnarray}
\dot \rho_\text{1}+3H(1+\omega_1)\rho_\text{1}=3\tilde H\rho_\text{1}(1+\omega_1)\,,\nonumber\\  
\dot \rho_\text{2}+3H(\rho_\text{2}+p_\text{2})=-3\tilde H\rho_\text{1}(1+\omega_1)\,,\label{EoS2}
\end{eqnarray}
according with (\ref{CL}). It is understood that the generic expression for the coupling must be derived in terms of $H\,,\rho_1\,,\rho_2$ only, as we will see in the specific examples. One possibility is always given by
\begin{equation}
3\tilde H\rho_\text{1}(1+\omega_1)=\left[3H\rho_1-\sqrt{3\kappa^2}\rho_1^{3/2}\right](1+\omega_1)\,,
\label{cap1}
\end{equation}
such that when the contribute of the fluid two disappears,
$H\simeq H_1=\sqrt{\kappa^2\rho_1/3}$ and the coupling vanishes, recovering the standard cosmology (for example, if $\omega_1=0$, the matter era takes place as in the Standard Model).

Finally, the Friedmann equations lead to
\begin{eqnarray}
\frac{3}{\kappa^2}\left(\tilde H^2+2 H_\text{1} \tilde H\right)=\rho_\text{2}\,,\nonumber\\
-\frac{1}{\kappa^2}\left(2\dot{\tilde H}+3\tilde H^2+6 H_\text{1}\tilde H\right)=p_\text{2}\,,\label{F2}
\end{eqnarray}
and the explicit contribute of the fluid one disappears. The assumption that (\ref{rhodm}) continues to be valid permits to cancel the contribute of the fluid one in the Friedmann equations, leading to an easy mathematical treatment of the model. The results that may be aquired in this way can be interesting, since the coupling between the fluids avoids some problem of standard cosmology like the coincidence problem (the Equations of State of the fluids are not independent), and also gives the possibility to analyze several cosmological solutions
by changing the aboundance of matter/dark energy with respect to the case of decoupled fluids. 

Let us investigate some solutions for accelerated universe by starting from this formalism.

\section{Dark energy solutions}

In order to derive the solutions able to describe an accelerated expansion, it is useful to introduce a general notation for the effective energy density and pressure in the Friedmann equations (\ref{EOMs}), namely
\begin{equation}
\rho_\text{eff}=\rho_\text{1}+\rho_\text{2}\,,\quad p_\text{eff}=p_\text{1}+p_\text{2}\,,
\end{equation}
such that the conservation law (\ref{CL}) reduces to
\begin{equation}
\dot\rho_\text{eff}+3 H(\rho_\text{eff}+p_\text{eff})=0\,.
\end{equation}
The simplest example of effective fluid is given by the perfect fluid, 
\begin{equation}
p_\text{eff}=\omega_\text{eff}\rho_\text{eff}\,,
\end{equation}
$\omega_\text{eff}$ being a (constant) effective Equation of State parameter. It follows from the first Friedmann equation (when $\omega_\text{eff}\neq -1$),
\begin{equation}
H(t)=\frac{2}{3(1+\omega_\text{eff})(t+t_0)}\,,
\quad
a(t)=a_0(t+t_0)^{\frac{2}{3(1+\omega_\text{eff})}}\,,\quad 
\rho_\text{eff}(t)=\rho_{\text{eff} (0)} a(t)^{-3(1+\omega_\text{eff})}\,,\quad\omega_\text{eff}\neq-1\,,\label{quint}
\end{equation}
$a_0\,,\rho_{\text{eff} (0)}\,,t_0$ being integration constants eventually related to each other.
Furthermore, when $\omega_\text{eff}=-1$, we get
\begin{equation}
H=\sqrt{\frac{\kappa^2\rho_{\text{eff}(0)}}{3}}\,,
\quad a(t)=a_0 \text{e}^{\sqrt{\kappa^2\rho_{\text{eff}(0)}/3} t}\,,
\quad\rho_\text{eff}=\rho_{\text{eff}(0)}\,,
\quad\omega_\text{eff}=-1\,.
\end{equation}
The strong energy condition (SEC) is violated for 
$\omega_\text{eff}<-1/3$, and $\ddot a/a=H^2+\dot H>0$, such that we obtain an acceleration.
In expanding universe ($H>0$), if $-1<\omega_\text{eff}<-1/3$ (quintessence fluid), the constant $t_0$ must be positive and it is usually set as $t_0=0$. On the other side, if $\omega_\text{eff}<-1$ (phantom fluid), in order to have an expansion we must put $t_0<0$ and $t<-t_0$. Thus, the Hubble parameter can be rewritten as
\begin{equation}
H(t)=-\frac{2}{3(1+\omega_\text{eff})(t_0-t)}\,,\quad t<t_0\,,\quad \omega_\text{eff}<-1\,,\label{phantom}
\end{equation}
where we have shifted $t_0\rightarrow-t_0>0$. In such a case, when $t$ is close to $t_0$,
the Hubble parameter, the Ricci scalar $R=12H^2+6\dot H$, the effective energy density and the scale factor diverge and the Big Rip scenario occurs~\cite{Caldwell}. In the next subesections, we will take in consideration this kind of solutions (in particular, the quintessence solution is related to the universe today with matter and dark energy), but
generally speaking the effective energy density of the universe can be represented by a non-perfect fluid.
The most important case in this sense is the $\Lambda$CDM Model, where the dark energy producing acceleration is not coupled with the matter and it is given by a perfect fluid whith EoS parameter $\omega=-1$.  In $\Lambda$CDM Model the first Friedmann equation reads
\begin{equation}
H(t)=\sqrt{\frac{\kappa^2}{3}}\sqrt{\frac{\rho_{\text{m(0)}}}{a(t)^3}+\frac{\Lambda}{\kappa^2}}\,,
\end{equation}
where $\rho_{\text{m(0)}}$ is the energy density of matter at $a(t)=1$ (usually one normalizes the scale factor $a(t)=1$ at the present epoch) and $\Lambda/\kappa^2$ is the constant energy density of dark energy. By solving this equation with respect to $a(t)$, one finds 
\begin{equation}
a(t)=\left(\frac{\kappa^2\rho_{m(0)}}{\Lambda}\right)^{1/3}\sinh\left[\sqrt{\frac{3}{4}\Lambda}t\right]^{2/3}\,,\quad\rho_{m(0)}\,,\Lambda\neq 0\,.
\end{equation}
The asymptotic limits of $a(t)$ are
\begin{equation}
a(t\rightarrow 0^+)\simeq
\left(\frac{3\kappa^2\rho_{m(0)}}{4}\right)^{1/3} t^{2/3}\,,\quad
a(t\rightarrow+\infty)\simeq \left(\frac{\kappa^2\rho_{m(0)}}{\Lambda}\right)^{1/3}\text{e}^{\sqrt{\frac{\Lambda}{3}}t}\,,\label{limits}
\end{equation}
and one recovers the results of matter and dark energy dominated eras, respectively. The Hubble parameter reads
\begin{equation}
H(t)=\sqrt{\frac{\Lambda}{3}}\coth\left[\sqrt{\frac{3\Lambda}{4}}t\right]\,.\label{Hl}
\end{equation}
By using the formalism presented in the second section, we would now to give a different picture of the universe contents which leads to the same solution of $\Lambda$CDM model, namely we will consider the coupling between the fluid one (\ref{fluidone}), which will be finally identified with matter, and the dark fluid two as in (\ref{EoS2}). By 
decomposing $H$ as in (\ref{tH}),
we get
\begin{equation}
\tilde H(t)=\sqrt{\frac{\Lambda}{3}}\coth\left[\sqrt{\frac{3\Lambda}{4}}t\right]-\frac{2}{3t(1+\omega_1)}\,.
\end{equation}
To solve the system (\ref{F2}) with respect to (\ref{Hl}) it is necessary a dark fluid with
\begin{equation}
\rho_2=\frac{\Lambda \coth\left[\sqrt{\frac{3\Lambda}{4}}t\right]^2}{\kappa^2}-\frac{4}{3\kappa^2t^2(1+\omega_1)^2}\,,\quad
p_2=-\frac{4\omega_1+3t^2\Lambda(1+\omega_1)^2}{3t^2\kappa^2(1+\omega_1)^2}\,,
\end{equation}
such that
\begin{equation}
\omega_2\equiv\frac{p_2}{\rho_2}=
\frac{-4\omega_1-3t^2\Lambda(1+\omega_1)^2}{-4+3t^2\Lambda(1+\omega_1)^2\coth\left[\sqrt{\frac{3\Lambda}{4}}t\right]^2}\,.\label{ooo}
\end{equation}
This expression for the Eos parameter induces a viscosity in the Equation of State of the fluid two, since the dependence on $H$ is manifest, namely
\begin{equation}
p_2=\omega_1\rho_2-\frac{\Lambda}{\kappa^2}-\frac{3H^2}{\kappa^2}\omega_1\,.
\end{equation}
Finally, the coupling between the two fluids in (\ref{EoS2}) is given by (\ref{cap1}). 

\begin{figure}[-!h]
\begin{center}
\includegraphics[angle=0, width=0.55\textwidth]{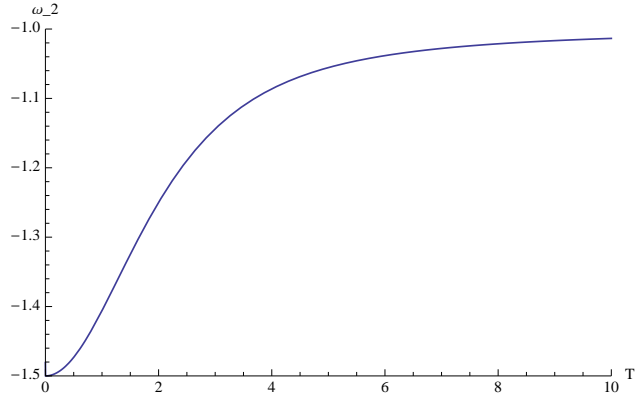}
\end{center}
\caption{Plot of $\omega_{2}$ as a function of $T=\sqrt{\Lambda}t$ in the case of $\omega_1=0$ in the coupled fluids model.\label{Fig1}}
\end{figure}

Let us take $\omega_1=0$, namely the standard matter case for the fluid one. 
The EoS parameter (\ref{ooo}) of the fluid two results to be
\begin{equation}
\omega_2=
-\frac{3t^2\Lambda}{4-3t^2\Lambda\coth\left[\sqrt{\frac{3\Lambda}{4}}t\right]^2}\,.
\end{equation}
In Fig.~(\ref{Fig1}) the graphic of $\omega_2$ as a function of $T=\sqrt{\Lambda}t$ is depicted.
We can see that for $t\rightarrow 0^+$ it tends to zero, while for $t\rightarrow+\infty$ it asymptotically tends to $-1$, namely the fluid two is a phantom fluid and never crosses the line of the phantom divide. The two limits correspond to the matter dominated universe and to the de Sitter universe whose scale factors are given by (\ref{limits}).

\begin{figure}[-!h]
\begin{center}
\includegraphics[angle=0, width=0.55\textwidth]{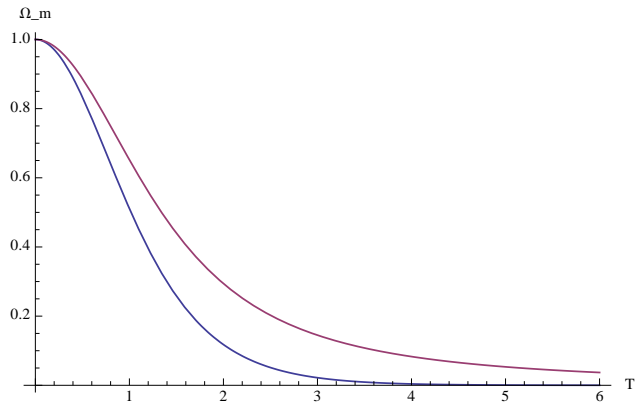}
\end{center}
\caption{Plot of $\Omega_{m}$ as a function of $T=\sqrt{\Lambda}t$ for the $\Lambda$CDM Model (blue line) and for the coupled fluids model
with $\omega_1=0$ (pink line).\label{Fig2}}
\end{figure}

The ratio between matter energy density and total effective energy density in the $\Lambda$CDM Model is
\begin{equation}
\Omega_m\equiv\frac{\rho_m}{\rho_\text{eff}}=\frac{\rho_\text{m(0)}a(t)^{-3}}{\rho_\text{m(0)}a(t)^{-3}+\frac{\Lambda}{\kappa^2}}=
\frac{1}{1+\sinh\left(\sqrt{\frac{3\Lambda}{4}}t\right)^2}\,,\quad\text{($\Lambda$CDM)}
\end{equation}
while for our coupled fluids model we get
\begin{equation}
\Omega_m\equiv\frac{\rho_m}{\rho_\text{eff}}=\frac{H_1(t)^2}{H(t)^2}=
\frac{4\tanh\left[\sqrt{\frac{3\Lambda}{4}}t\right]^2}{3t^2\Lambda}\,.
\end{equation}
Here, $\Omega_m$ is the cosmological parameter of matter and
in the last expression we have put $\rho_m=\rho_1$ with $\omega_1=0$. In Fig.~(\ref{Fig2}) we show the graphics of $\Omega_m$ in the case of $\Lambda$CDM Model and in the case of the coupled fluids model under investigation. 

For $t\rightarrow 0^+$, $\rho_m/\rho_\text{eff}\rightarrow 1$ in the both cases and the matter is dominant: in the coupled fluids model the contribute of the dark fluid two is avoided and the matter dominated era is reproduced in the same way of the Standard Model. On the other side, when $t$ grows up, the contribute of matter decreases in two different way: for the $\Lambda$CDM Model we have $\Omega_m(t\rightarrow+\infty)\sim\exp[-t]^2$, while for the coupled fluids model, $\Omega_m(t\rightarrow+\infty)\sim t^{-2}$. It means that in the coupled fluids model, even if   matter finally
disappears in the de Sitter universe, its energy density decreases slowlier with respect to the case of $\Lambda$CDM Model, as it is clear from the Fig.~(\ref{Fig2}).
For example, in the universe today, at the time $t=t_0$, the ratio between matter and effective energy density predicted by $\Lambda$CDM Model is
\begin{equation}
\Omega_m(t_0)=\frac{\Lambda}{3 H(t_0)^2}\frac{1}{\sinh\left[\sqrt{\frac{3\Lambda}{4}}t_0\right]^2}\,,\quad\text{($\Lambda$CDM)}
\end{equation} 
while in the coupled fluids model reads
\begin{equation}
\Omega_m(t_0)=\frac{4}{9t_0^2H(t_0)^2}\,.
\end{equation} 
Since $\Lambda/H(t_0)^2\simeq 3$ and in the $\Lambda$CDM Model $\rho_m/\rho_\text{eff}\simeq 0.32$, we can estimate $\sqrt{\Lambda}t_0$ as 
$\sqrt{\Lambda}t_0\simeq 1.54$. As a consequence, in the case of the coupled fluids model, we obtain for the universe today $\Omega_m(t_0)\simeq 0.56$, namely the amount of dark fluid results to be smaller with respect to the case of $\Lambda$CDM Model.

Thus, in a coupled fluids model like the one analyzed in this paper, it is possible to recover the dynamics of $\Lambda$CDM Model: the scale factor, the Hubble parameter (and more in general all the cosmographic parameters) and the effective EoS parameter of the universe can be reproduced in the same way, according with the cosmological data.
When matter (here, the fluid one with $\omega_1=0$) is dominant, the coupling (\ref{cap1}) vanishes and the behaviour of matter obviously is the same of the Standard Model. The future evolution of the model asymtotically tends to the de Sitter epoch where a dark fluid with $\omega_2=-1$ is dominant: the contribute of matter disappears and the energy density of the fluid two can be identified with the Cosmological Constant. The differences between the model under investigation and the $\Lambda$CDM Model emerge in the intermediate epoch, when the coupling between the Equations of State of the two fluids becomes important. The dark fluid two remains in the phantom region, but its EoS parameter is much smaller than $-1$.
The amounts of matter and dark fluid depend on each other and the coincidence problem is removed. Furthermore, due to the coupling, the energy density of the dark fluid results to be smaller than the one necessary in the $\Lambda$CDM Model to obtain the accelerated solution of the universe today. Since the observed (baryonic) matter composes only the $5\%$ of the universe, this model necessary brings to a larger amount of dark matter.  

In the nexts subsections, we will analyze the cases of effective quintessence and phantom perfect fluids for accelerating universe.

\subsection{Quintessence solutions}

Let us consider the quintessence solution for expanding universe in our two coupled fluids model. The total effective EoS parameter is such that $-1<\omega_\text{eff}<-1/3$ and the Hubble parameter can be written as in Eq.~(\ref{quint}) with $t_0=0$. By decomposing the Hubble parameter as in (\ref{tH}), where $H_\text{1}$ is given by (\ref{HDM}), one has
\begin{equation}
\tilde H=-\frac{2(\omega_\text{eff}-\omega_1)}{3t(1+\omega_\text{eff})(1+\omega_1)}\,,
\end{equation}
and the Friedmann equations (\ref{F2}) lead to
\begin{equation}
\rho_\text{2}=\frac{4(\omega_\text{1}-\omega_\text{eff})(2+\omega_\text{eff}+\omega_1)}{3\kappa^2 t^2(1+\omega_\text{eff})^2(1+\omega_1)^2}\,,\quad 
p_\text{2}=\frac{4(\omega_\text{eff}-\omega_1)(1-\omega_1\omega_\text{eff})}{3\kappa^2 t^2(1+\omega_\text{eff})^2(1+\omega_1)^2}\,,
\end{equation}
such that the EoS paramter of the fluid two finally reads
\begin{equation}
\omega_\text{2}\equiv \frac{p_\text{2}}{\rho_\text{2}}=\frac{\omega_1\omega_\text{eff}-1}{2+\omega_\text{eff}+\omega_1}\,.\label{omegaf2}
\end{equation}
The coupling between the two fluids in (\ref{EoS2}) may be expressed in terms of $\rho_1\,,\rho_2$ as
\begin{equation}
3\tilde H\rho_\text{1}(1+\omega_1)=
\sqrt{\frac{\omega_1-\omega_\text{eff}}{2+\omega_\text{eff}+\omega_1}}
\sqrt{3\kappa^2\rho_2}\rho_1(1+\omega_1)\,.
\end{equation}
Let us consider the case of $\omega_1=0$, namely the fluid one  is identified with the matter. The quintessence universe results to have a constant ratio between matter energy density and effective energy density, and matter energy density and dark fluid energy density,
\begin{equation}
\frac{\rho_1}{\rho_\text{eff}}\equiv\frac{(1+\omega_\text{eff})^2}{(1+\omega_1)^2}=
(1+\omega_\text{eff})^2\,,\quad
\frac{\rho_\text{1}}{\rho_\text{2}}\equiv-\frac{(1+\omega_\text{eff})^2}{(\omega_\text{eff}-\omega_1)(2+\omega_\text{eff}+\omega_1)}=
-\frac{(1+\omega_\text{eff})^2}{\omega_\text{eff}(2+\omega_\text{eff})}\,.
\label{quintratio}
\end{equation}
If $\omega_\text{eff}=-0.68$, namely the value of the universe today, we get
\begin{equation}
\omega_2\simeq-0.75\,,\quad\frac{\rho_1}{\rho_\text{eff}}\simeq 0.10\,,\quad\frac{\rho_1}{\rho_2}\simeq 0.11\,.
\end{equation}
The dark fluid which supports the quintessence solution results to be also a quintessence fluid.
The model can reproduce the expansion of the universe today mantaining constant the ratio between matter and dark fluid, and therefore predicting a different future evolution with respect to the $\Lambda$CDM Model. In fact, here the universe will remain in an eternal quintessence phase. 
Due to the coupling, the coincidence problem is removed and the dark fluid which brings to the acceleration has an EoS parameter much larger than the one of decoupled dark energy (which must be very close to minus one to produce the same expansion today). Furthermore,
the total amount of matter results to be smaller than the one in $\Lambda$CDM Model (where $\rho_m/\rho_\text{eff}\simeq 0.32$), rendering also smaller the contribute of dark matter (the observed baryonic matter being at $5\%$).

\subsection{Phantom solutions}

In order to study the phantom solution (\ref{phantom}) in our two coupled fluids model, it is convenient to shift $t\rightarrow -(t_0-t)$ in $H_1$ of Eq.~(\ref{HDM}), being $t_0$ and integration constant of the solution. The formalism remains still valid (since $\rho_1\sim  a_1(t)^{3(\omega_1+1)}$) and, due to the contribution of the second dark fluid, the solution is again for expanding universe, despite to the fact that $H_1<0$ when $-1<\omega_1$ . The introduction of the integration constant $t_0$ in $H_1$ appears to be quite natural since it is also present in $H$. In this case,
by decomposing the Hubble parameter as in (\ref{tH}), we get
\begin{equation}
\tilde H(t)=\frac{2(\omega_\text{eff}-\omega_1)}{3(t_0-t)(1+\omega_\text{eff})(1+\omega_1)}\,,
\end{equation}
and from the system (\ref{F2}) one easily derives the energy density and pressure of the fluid two,
\begin{equation}
\rho_\text{2}=\frac{4(\omega_\text{1}-\omega_\text{eff})(2+\omega_\text{eff}+\omega_1)}{3\kappa^2 (t_0-t)^2(1+\omega_\text{eff})^2(1+\omega_1)^2}\,,\quad 
p_\text{2}=\frac{4(\omega_\text{eff}-\omega_1)(1-\omega_1\omega_\text{eff})}{3\kappa^2 (t_0-t)^2(1+\omega_\text{eff})^2(1+\omega_1)^2}\,.
\end{equation}
Here, we note that, in order to have a positive energy density for that fluid, one must require
\begin{equation}
-2-\omega_1<\omega_\text{eff}<-1\,.\label{cond}
\end{equation}
The EoS parameter of the fluid two returns to be (\ref{omegaf2}) and
the coupling with the fluid one may be now expressed as 
\begin{equation}
3\tilde H\rho_\text{1}(1+\omega_1)=-
\sqrt{\frac{\omega_1-\omega_\text{eff}}{2+\omega_\text{eff}+\omega_1}}
\sqrt{3\kappa^2\rho_2}\rho_1(1+\omega_1)\,.
\end{equation}
The phantom universe obtained in this way has the same constant ratio between the fluids of the quintessence case (\ref{quintratio}). Let us take $\omega_1=0$, such that the first fluid can be identified with matter. In order to satisfy the condition (\ref{cond}), we need 
\begin{equation}
-2<\omega_\text{eff}<-1\,,\quad\omega_1=0\,,
\end{equation}
such that the ratio between matter and effective energy density correctly results smaller than one. 
As a consequence, the dark fluid is phantom since
\begin{equation}
-\infty<\omega_2<-1\,.
\end{equation}
The interesting point is that the ratio between the matter and the dark fluid two can be larger than one when
\begin{equation}
1<\frac{\rho_1}{\rho_2}\,,\quad\omega_1=0\,,\quad\-2<\omega_\text{eff}<-1-\frac{1}{\sqrt{2}}\simeq-1.7\,.
\end{equation}
It means that, due to the coupling between the fluids, we can obtain an acceleration even if the contribute of matter is larger than the one of the dark fluid. This fact is quite interesting. The acceleration comes from a dark fluid (with $\omega_2<-1$), but it has to be not  necessarily dominant.
We finally remark that this kind of accelerated solution is phantom, and the universe will end with a Big Rip at the time $t=t_0$.

\subsection{Other accelerated solutions}

Due to the presence of the coupling between the fluids, other accelerated solutions are allowed.
An example is given by the finite-future time singularity solutions, which are a generalization of (\ref{phantom}), namely
\begin{equation}
H(t)=\frac{h_0}{(t_{0}-t)^{\beta}}\,,\quad t<t_0\,,\beta\neq 0\,.
\label{Hsingular}
\end{equation}
Here, $h_0$, and $t_{0}$ are positive constants,
$\beta$ is a generic parameter which describes the type of singularity and $t_0$ is the finite-time for which singularity occurs (in the future). At that time, the Hubble parameter or its derivative diverge.
The strongest singularities are obtained for $1<\beta$, where Hubble parameter, Ricci scalar and scale factor diverge~\cite{Rip1}--\cite{Rip7}, or for $0<\beta<1$ (sudden), where Hubble parameter and Ricci scalar diverge~\cite{sudden, suddenOd}. For $-1<\beta<0$ only Ricci scalar diverges~\cite{IIIOd} and for $\beta<-1$ some derivetives of Hubble parameter become singular~\cite{classificationSingularities}. For $\beta=1$ we recover the Big Rip case (\ref{phantom}).
This kind of solutions violate the SEC (at least near to the singularity), namely bring the universe to an acceleration and have been often studied as possible future scenarios for the dark energy epoch.

By decomposing the Hubble parameter as in (\ref{tH}) with respect to (\ref{Hsingular}), one obtains
\begin{equation}
\tilde H=\frac{3h_0(1+\omega_1)+2(t_0-t)^{\beta-1}}{3(t_0-t)^\beta(1+\omega_1)}
\,.
\end{equation}
Here , we took again $t\rightarrow -(t_0-t)$ in (\ref{HDM}).
System (\ref{F2}) can be solved as
\begin{equation}
\rho_\text{2}=
\frac{9h_0^2(1+\omega_1)^2-4(t_0-t)^{2\beta-2}}{3\kappa^2(t_0-t)^{2\beta}(1+\omega_1)^2}
\,,\quad 
p_\text{2}=
-\frac{3h_0(1+\omega_1)^2\left(3h_0+2(t_0-t)^{\beta-1}\beta\right)+4\omega_1(t_0-t)^{2\beta-2}}{3\kappa^2(t_0-t)^{2\beta}(1+\omega_1)^2}\,.
\end{equation}
The positivity of the energy density of this fluid must be carefully analyzed and depends on $\beta$. If $\beta>1$, the energy density is positive at least near to the singularity and diverges at $t_0$ with the Hubble parameter and the Ricci scalar. On the other side, if $\beta<-1$, the energy density becomes negative making the fluid unphysical near to the singularity. For this reason, for our purpose we will consider only the case
\begin{equation}
1<\beta\,.
\end{equation}
The EoS parameter of fluid two reads
\begin{equation}
\omega_2=-1
-\frac{6h_0\beta(1+\omega_1)^2(t_0-t)^{\beta-1}+4(\omega_1+1)(t_0-t)^{2\beta-2}}
{9h_0^2(1+\omega_1)^2-4(t_0-t)^{2\beta-2}}\,,
\end{equation}
namely
\begin{equation}
p_2=-\rho_2
-\left(\frac{
6h_0^\frac{-1}{\beta}\beta H^{\frac{1+\beta}{\beta}} +4(\omega_1+1)^{-1}H^{\frac{2}{\beta}}h_0^{-\frac{2}{\beta}}}
{3\kappa^2}\right)\,,
\end{equation}
such that a viscosity term appears in the Equation of State of such fluid.
The coupling between the two fluids is given by (\ref{cap1}), but also other expressions can be found, like for example
\begin{equation}
3\tilde H\rho_\text{1}(1+\omega_1)=
\left[3H(1+\omega_1)+2\left(\frac{h_0}{H}\right)^{1/\beta}\right]\rho_1\,.\label{cccp}
\end{equation}
Despite to the fact that on the solution (\ref{Hsingular}) this expression obviously coincides with (\ref{cap1}), the generic expression of the coupling between the two fluids determines the behaviour of the model with respect to other (possible) solutions. For example, with the coupling term (\ref{rhodm}) in the Equations of State of the two fluids, we can recover the matter solution (with $\omega_1=0$) when the fluid two disappears , but with the coupling term above, the fluid two never disappears (if $\rho_1\neq 0$) and the matter solution cannot be found.

The ratio between the energy density of the fluid one and the total effective energy density and the ratio between the two fluids energy densities read
\begin{equation}
\frac{\rho_1}{\rho_\text{eff}}=\frac{4(t_0-t)^{2(\beta-1)}}{9h_0^2(1+\omega_1)^2}\,,
\quad
\frac{\rho_1}{\rho_2}=\frac{4(t_0-t)^{2(\beta-1)}}{9h_0^2 (1+\omega_1)^2-4(t_0-t)^{2(\beta-1)}}\,.
\end{equation}
Thus, near to the singularity, by considering $\beta>1$, the fluid one (which may be identified with matter for $\omega_1=0$) tends to vanish with respect to the dark fluid producing the singularity. For $\beta=1$ we recover the constant ratios of phantom case.\\
\\
An other interesting class of solutions which provide acceleration is given by the bounce solutions. The bounce cosmology has been suggested many years ago as an alternative scenario to the Big Bang theory. In the presence of a bounce, a cosmological contraction is followed by an expansion at a finite time and the universe emerges from the bounce instead to the initial singularity of the Big Bang  (see Ref.~\cite{Novello} for a review).
A simple example of bounce solution can be easily derived from the Big Rip solution (\ref{phantom}),
\begin{equation}
a(t)=\alpha(t-t_0)^{2n}\,,\quad H (t)=\frac{2n}{(t-t_0)}\,,\quad n=1,2,3...
\label{pow}
\end{equation}
Here, $\alpha$ is a positive (dimensional) constant and $n$ is a positive natural number from which depends the feature of the bouncing. Moreover, $t_0>0$ is the fixed bounce time. When $t<t_0$, we have  a contraction with negative Hubble parameter, at $t=t_0$ we have the bounce, and when $t>t_0$ the universe expands with positive Hubble parameter. Note that this kind of solution leads to an acceleration before and after the bounce (see also Ref.~\cite{miobounce}).
We mention this solution since it is easy to reproduce it for our coupled fluids model
by starting from the finite-future time singularity case above presented, making the substitutions $h_0\rightarrow -2n$ and $\beta=1$ in all the formulas. For the bounce solution (\ref{pow}), the energy density and pressure of fluid two read
\begin{equation}
\rho_2=\frac{4\left(9n^2(1+\omega_1)^2-1\right)}{3\kappa^2(t-t_0)^2(1+\omega_1)^2}\,,\quad
p_2=-\frac{4\left(3n(1+\omega_1)-1\right)\left(3n(1+\omega_1)-\omega_1\right)}{3\kappa^2(t-t_0)^2(1+\omega_1)^2}\,,
\end{equation}
such that the energy density can be positive defined if $9n^2(1+\omega_1)^2>1$. The EoS parameter of this fluid is constant and reads
\begin{equation}
\omega_2=\frac{\omega_1-3n(1+\omega_1)}{1+3n(1+\omega_1)^2}\,,
\end{equation}
such that the fluid is in the quintessence region (for example, for $\omega_1=0$, $\omega_2=-3n/(1+3n)$, $-1<\omega_2<-3/4$). The ratio between the energy density of the first fluid and the total effective energy density and the ratio between the energy densities of the two fluids   read
\begin{equation}
\frac{\rho_1}{\rho_\text{eff}}=\frac{1}{9n^2(1+\omega_1)^2}\,,
\quad
\frac{\rho_1}{\rho_2}=\frac{1}{9n^2(1+\omega_1)^2-1}\,,
\end{equation}
namely they are constant.

\section{Other applications}

We conclude the work by presenting an extension of our model to other types of fluid one, with Equation of State different to (\ref{fluidone}), namely with $\omega_1$ not a constant. We will analyze a simple example of non perfect fluid, whose Equation of State reads
\begin{equation}
p_1=\omega_1(\rho_1)\rho_1\,,\quad\omega_1(\rho_{1})=A_{0}\rho_{1}^{\alpha-1}-1\label{omega}\,,
\end{equation}
where $A_{0}>0$ and $\alpha>1$ are assumed to be constants.  We consider the two fluids model (\ref{EOMs}) again. When the contribute of the fluid two vanishes, the energy conservation law leads to
\begin{equation}
\rho_{1}=\left[(\alpha-1)\left(3A_{0}\ln \frac{a_1(t)}{a_{0}}\right)\right]^{\frac{1}{1-\alpha}}\,,\label{above}
\end{equation}
where $a_{0}$ is a positive scale parameter. Note that the fluid energy density is positive defined due to the assumptions on $A_0$ and $\alpha$. We rewrite (\ref{above}) in the following way,
\begin{equation}
\rho_{1}=\frac{H_{0}^{2}}{\kappa^2}\left[\ln \frac{a_1(t)}{a_{0}}\right]^{\frac{1}{1-\alpha}}\,,
\quad
H_0^2=\kappa^2[3A_{0}(\alpha-1)]^{\frac{1}{1-\alpha}}\,.
\label{Lorenzo}
\end{equation}
From Friedmann equations we get
\begin{equation}
H_1(t)=\frac{Z_0}{t^{1/(2\alpha-1)}}\,, 
\quad
a_1(t)=a_{0}\,\text{Exp}\left\{Z_0 t^{2(\alpha-1)/(2\alpha-1)}\right\}\label{zirp}\,,
\end{equation}
where
\begin{equation}
Z_0 =6^{\frac{2-2\alpha}{2\alpha-1}}\left[\dfrac{(2\alpha-1)(\sqrt{3}H_{0})}{\alpha-1}\right]^{2(\alpha-1)/(2\alpha-1)}\,.
\end{equation}
For large values of $\alpha$, the fluid energy density tends to $\rho_1\simeq H_{0}^{2}/\kappa^2$, the EoS parameter reads $\omega(\rho_{1})\simeq -1$ and the scale factor is $a_1(t)\simeq a_{0}\mathrm{e}^{H_{0}t/3}$ (de Sitter universe), but in general $\omega_1>-1$. We can see now how a second fluid coupled with this can change the dynamics of the model. We may look for the phantom solution (\ref{phantom}) again, where the total effective EoS parameter $\omega_\text{eff}$ is smaller than minus one. In analogy with the `matter' case, we shift $t\rightarrow (t_0-t)$ in (\ref{zirp}), obtaining a scale factor which is still a solution of the Friedmann equations for the fluid (\ref{omega}), since in principle the solutions of Friedmann equations are 
$a (t)\sim \exp[\pm t^{2(\alpha-1)/(2\alpha-1)}]$.

By solving the system (\ref{F2}) with $H_1$ given by (\ref{zirp}), one obtains for energy density and pressure of the fluid two
\begin{eqnarray}
\rho_2&=&\frac{4(t_0-t)^\frac{2-4\alpha}{2\alpha-1}-9(t_0-t)^\frac{-2}{2\alpha-1}Z_0^2(1+\omega_\text{eff})}{3\kappa^2(1+\omega_\text{eff})^2}\,,
\nonumber\\
p_2&=&\frac{9Z_0^2(1+\omega_\text{eff})^2(t_0-t)^\frac{2}{1-2\alpha}
+6(1+\omega_\text{eff})^2 Z_0 (t_0-t)^{\frac{2\alpha}{1-2\alpha}}(2\alpha-1)^{-1}+4\omega_\text{eff}(t_0-t)^{-2}
}{3\kappa^2(1+\omega_\text{eff})^2}\,.\nonumber\\\label{dudu}
\end{eqnarray}
The related EoS parameter is
\begin{equation}
\omega_2=\frac{9Z_0^2(1+\omega_\text{eff})^2(t_0-t)^\frac{2}{1-2\alpha}
+6(1+\omega_\text{eff})^2 Z_0 (t_0-t)^{\frac{2\alpha}{1-2\alpha}}(2\alpha-1)^{-1}+4\omega_\text{eff}(t_0-t)^{-2}}{4(t_0-t)^\frac{2-4\alpha}{2\alpha-1}-9(t_0-t)^\frac{-2}{2\alpha-1}Z_0^2(1+\omega_\text{eff})}\,,
\end{equation}
namely a viscosity term depending on the velocity ($H$) appears in the Equation of State of such a fluid.
Finally, the ratio between the two fluids of the model is given by
\begin{equation}
\frac{\rho_1}{\rho_2}=\frac{9 Z_0^2(1+\omega_\text{eff})^2}{4(t_0-t)^{\frac{-4(\alpha-1)}{(2\alpha-1)}}
-9Z_0^2(1+\omega_\text{eff})^2}\,.
\end{equation}
The positivity of this result depends on the positivity of $\rho_2$ in (\ref{dudu}). From here we can see that near to the singularity, when the Hubble parameter diverges, the contribute of the fluid (\ref{omega}) disappears with respect to the one of the second fluid which in fact drives the phantom universe.

\section{Conclusions}

In this paper, we have analyzed a two coupled fluids model by investigating several solutions for accelerated universe in flat FRW space-time. The first fluid can reproduce the matter, and the second one is a dark fluid bringing the acceleration. Moreover, it is in general possible to recover also the standard matter era as a solution of the model. Beyond the removal of the coincidence problem and the simplicity of the mathematical treatment, we have seen in the specific examples how the coupling can change the description of the universe energy contents with respect to the cases of standard decoupled fluids. The model under investigation can reproduce the solution of $\Lambda$CDM Model with matter and dark energy and the quintessence solution that also may mimic the current expansion of the universe. In the first case, the dark energy of the universe results to be smaller than the one in $\Lambda$CDM Model, such that a larger amount of dark matter is necessary. In the second case, the quintessence universe where matter is coupled with a quintessence fluid, the effective EoS parameter $\omega_\text{eff}=-0.68$ of the universe today can be found by reducing the amount of dark matter. Also phantom, finite-future time singularity and bounce solutions (the last ones for inflation) have been investigated. It is interesting to observe that due to the coupling it is also possible to recover accelerated solutions with an amount of matter larger than the one of the dark fluid. In order to complete the work, a generalization of the model to the case of two fluids with non constant EoS parameter has been presented: a simple application has been analyzed as an example.

Other relevant works on inhomogeneous viscous fluids and the dark energy issue have been presented in Ref.~\cite{Ciappi}, in Refs.~\cite{uno}--\cite{seibis}, in Ref.~\cite{LittleRip} for viscous fluids in Little Rip cosmology, in Refs.~\cite{Carro, SuperJamil} for other fluid interactions and in Ref.~\cite{pp} for fluid perturbations in FRW universe.


\end{document}